\documentclass[runningheads]{llncs}
\usepackage{graphicx}
\begin{document}
\title{Collaborative Multi-agent Learning for MR Knee Articular Cartilage Segmentation}
\titlerunning{Collaborative Multi-agent Learning for MR Knee Cartilage Segmentation}
\author{Chaowei Tan\inst{1} \and 
        Zhennan Yan\inst{2} \and 
        Shaoting Zhang\inst{2} \and \\ 
        Kang Li\inst{1,3} \and 
        Dimitris N. Metaxas\inst{1}} 
%
\authorrunning{C. Tan et al.}
\institute{Department of Computer Science, Rutgers University, Piscataway, USA \and SenseTime Research \and Department of Orthopaedics, New Jersey Medical School, Rutgers University, Newark, USA}

\maketitle              
\begin{abstract}
The 3D morphology and quantitative assessment of knee articular cartilages (i.e., femoral, tibial, and patellar cartilage) in magnetic resonance (MR) imaging is of great importance for knee radiographic osteoarthritis (OA) diagnostic decision making. However, effective and efficient delineation of all the knee articular cartilages in large-sized and high-resolution 3D MR knee data is still an open challenge. In this paper, we propose a novel framework to solve the MR knee cartilage segmentation task. The key contribution is the adversarial learning based collaborative multi-agent segmentation network. In the proposed network, we use three parallel segmentation agents to label cartilages in their respective region of interest (ROI), and then fuse the three cartilages by a novel ROI-fusion layer. The collaborative learning is driven by an adversarial sub-network. The ROI-fusion layer not only fuses the individual cartilages from multiple agents, but also backpropagates the training loss from the adversarial sub-network to each agent to enable joint learning of shape and spatial constraints. Extensive evaluations are conducted on a dataset including hundreds of MR knee volumes with diverse populations, and the proposed method shows superior performance.
\end{abstract}

\section{Introduction}
\begin{figure}[!t]
  \centering
  \begin{minipage}[b]{0.3\linewidth}
    \centering
    \centerline{\includegraphics[width=\linewidth]{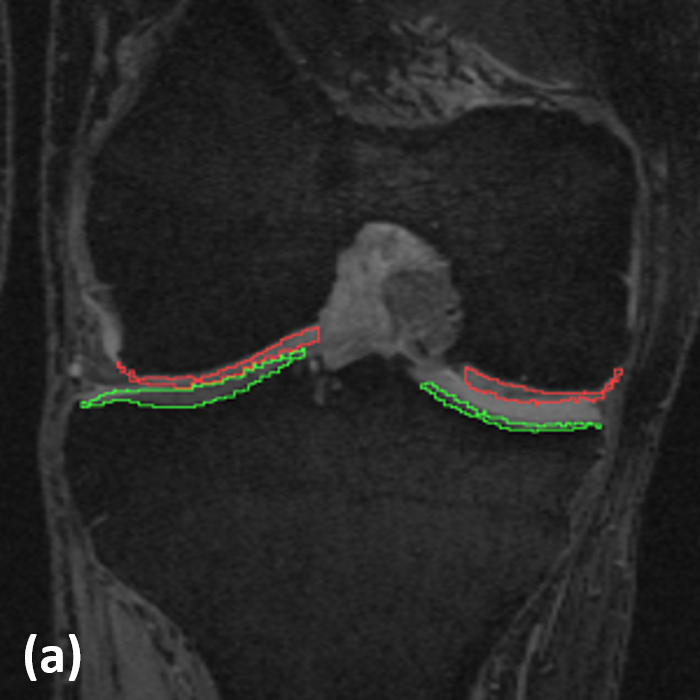}}
  \end{minipage}
  \hfil
  \begin{minipage}[b]{0.3\linewidth}
    \centering
    \centerline{\includegraphics[width=\linewidth]{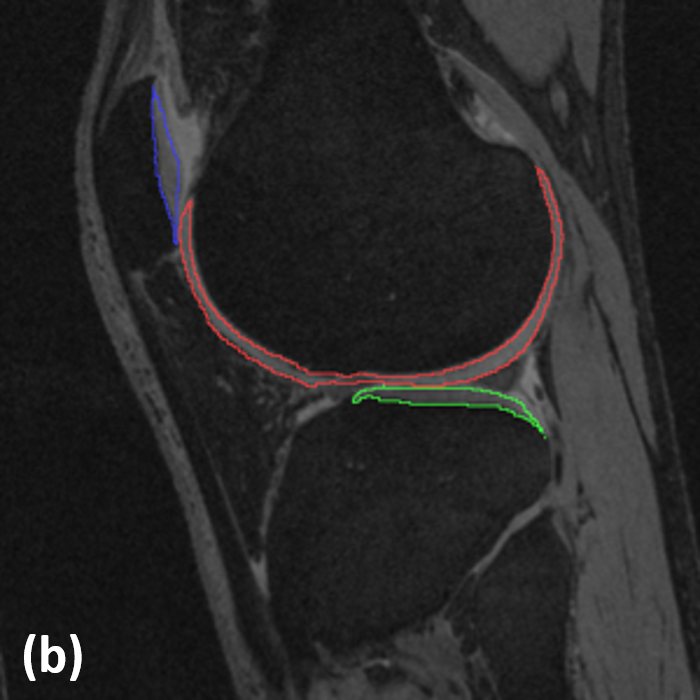}}
  \end{minipage}
  \hfil
  \begin{minipage}[b]{0.3\linewidth}
    \centering
    \centerline{\includegraphics[width=\linewidth]{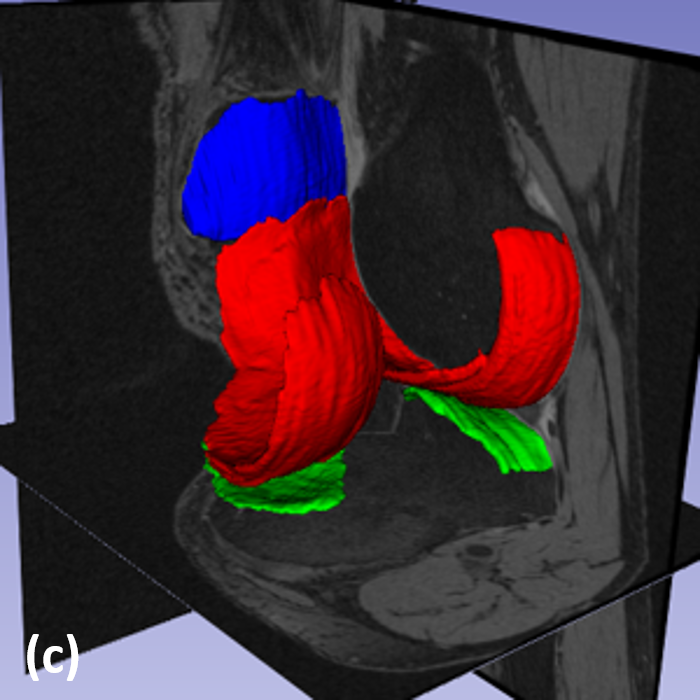}}
  \end{minipage}
  \caption{(a) and (b) show the coronal and sagittal slices of a 3D MR knee data. The red, green and blue contours indicate the femoral cartilage (FC), tibial cartilage (TC) and patellar cartilage (PC), respectively. (c) demonstrates the cartilage labels in 3D.}
  \label{fig:intro_anatomy_images}
\end{figure}

Osteoarthritis (OA) is the most common chronic health problem of human joints and the knee has the highest risk of developing OA in human lifetime. The knee articular cartilages (i.e., femoral, tibial, and patellar cartilage) are essential tissues for knee radiographic OA diagnosis. Eckstein et al.~\cite{eckstein2010quantitative} indicated that the cartilage morphology outcomes (e.g., cartilage thickness and surface area) by measuring 3D magnetic resonance (MR) data in knee joint can help to identify the symptomatic and structural severity of knee OA. Hunter et al.~\cite{hunter2011evolution} investigated the knee cartilage defects/losses by MR imaging as one important factor of knee OA. In order to capture the wide range and thin structure of cartilages in detail, MR data is usually in large size (millions of voxels) and high resolution. Fig.~\ref{fig:intro_anatomy_images} exhibits a 3D MR knee data from the  Osteoarthritis Initiative (OAI) database\footnote[1]{http://www.oai.ucsf.edu/}, which has high resolution ($0.365mm \times 0.365mm \times 0.7mm$) and large size ($384 \times 384 \times 160$). Effective and efficient segmentation of all articular cartilages in such high-resolution and large-sized data is challenging. Furthermore, the radiographic representations of cartilages may vary a lot in individuals with different age and pathology. Although the over-the-counter deep learning methods (e.g. VNet~\cite{milletari2016v}) have shown superior performances in many segmentation tasks, simply applying VNet to the MR knee data may have low accuracy and result in crash of training due to huge GPU memory consumption. Besides, the task of multi-cartilage classification suffers from severe class imbalance problem. Xu et al.~\cite{xu2018contextual} showed a contextual additive network focusing on the boost of memory efficiency for cartilage segmentation. The approach is based on small overlapping patches (a patch may only capture partial target) which may sacrifice certain accuracy. Some previous methods~\cite{he2019pelvic,tan2018deep} present multi-task networks. They introduce the distinctive boundary features of organ to improve accuracy. But the tissue of cartilage has very thin structure and its topology may change in degenerative conditions. Xu et al.~\cite{xu2018mutgan} segmented thin objects in 2D images through a myocardial infarction segmentation. Yet this 2D task-specific strategy may still suffer from the memory issue when applying for the 3D knee data.

\begin{figure}[!t]
    \centering
    \includegraphics[width=0.95\linewidth]{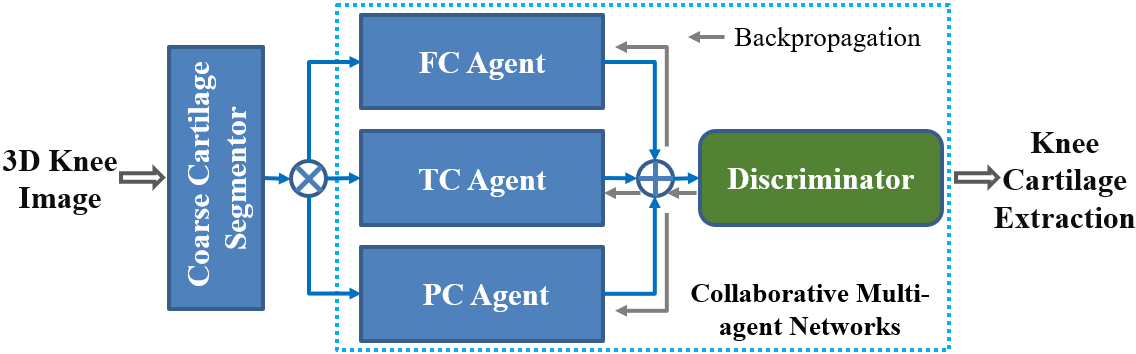}
    \caption{Flowchart of the collaborative multi-agent learning for cartilage segmentation.}
    \label{fig:framework}
\end{figure}

In this paper, we propose a novel segmentation framework with collaborative multi-agent learning (shown in Fig. \ref{fig:framework}) for the task of knee cartilage labeling in large-sized and high-resolution 3D MR data. Through region of interest (ROI) extraction, three high-resolution cartilage ROIs are fed into different segmentation agents. The multiple agents collaborate by the help of discriminator and produce cartilage labels at the end. The ROI-fusion layer not only fuses the individual cartilages from multiple agents for discriminator, but also backpropagates the training errors from the adversarial sub-network to each agent to enable joint learning of shape and spatial constraints. Such collaborative multi-agent framework can obtain fine-grained segmentation in each ROI and ensure the spatial constraints between different cartilages. It satisfies the limits of GPU resources and enables smooth training on the challenging data. The experimental results show that the proposed method can extract all cartilages accurately.

\begin{figure}[!t]
    \centering
    \includegraphics[width=0.95\linewidth]{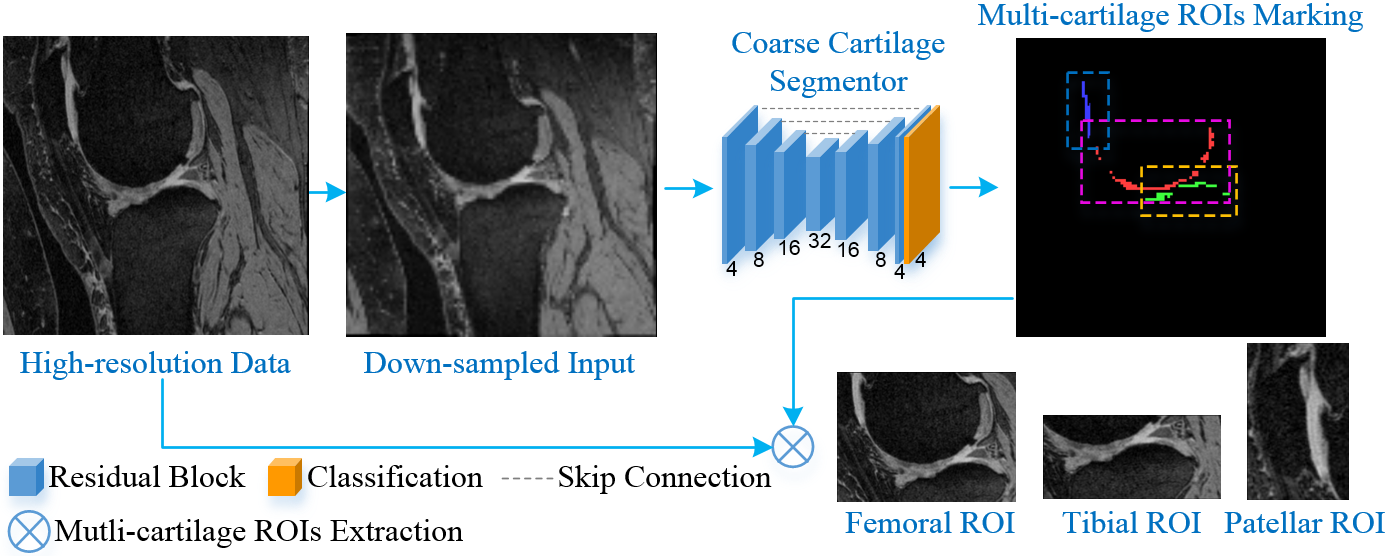}
    \caption{Overview of the multiple cartilage ROIs extraction (only show the sagittal view). The number of feature maps in the network is displayed under each block.}
    \label{fig:segmentor}
\end{figure}

\section{Methods} \label{sec:method_frame}
The overview of the proposed framework is shown in Fig.~\ref{fig:framework}. The coarse cartilage segmentor and ROI extraction (i.e., $\bigotimes$) steps aim to efficiently localize and extract three local regions of FC, TC and PC, and feed the ROIs to segmentation agents respectively. The blue dashed box shows the collaborative multi-agent cartilage segmentation module, which consists of three segmentation agents, one ROI-fusion layer (i.e., $\bigoplus$), and one joint-label discriminator.

\noindent \textbf{ROI extraction}. In order to initialize the collaborative multi-agent learning, we first extract the ROIs of three cartilages. As shown in Fig.~\ref{fig:segmentor}, by utilizing the location information of the multi-cartilage marks from the coarse segmentor, the image and label ROIs of FC, TC and PC are extracted from the original data. The segmentor's structure is like VNet~\cite{milletari2016v}, i.e., encoding-decoding. The encoding part contains 3 down-samplings (by convolutions of filter size 2 and stride 2) to obtain 3 different scales of feature maps. The decoding part has 3 up-samplings (by deconvolutions of filter size 2 and stride 2) to restore the scale of feature maps to reach the original input size. The blue block in this figure represents residual block followed by a down-sampling or up-sampling layer mentioned above when changing resolution. All the convolutional layers in the residual blocks have filter size 3, stride 1 and zero-padding 1. PReLU activation and batch normalization follow the convolutional and deconvolutional layers. The coarse cartilage segmentor is trained based on multi-class cross entropy loss $\ell_{mce}$ to obtain cartilage masks from the down-sampled MR data (e.g., $192 \times 192 \times 160$).

\begin{figure}[!t]
    \centering
    \includegraphics[width=\linewidth]{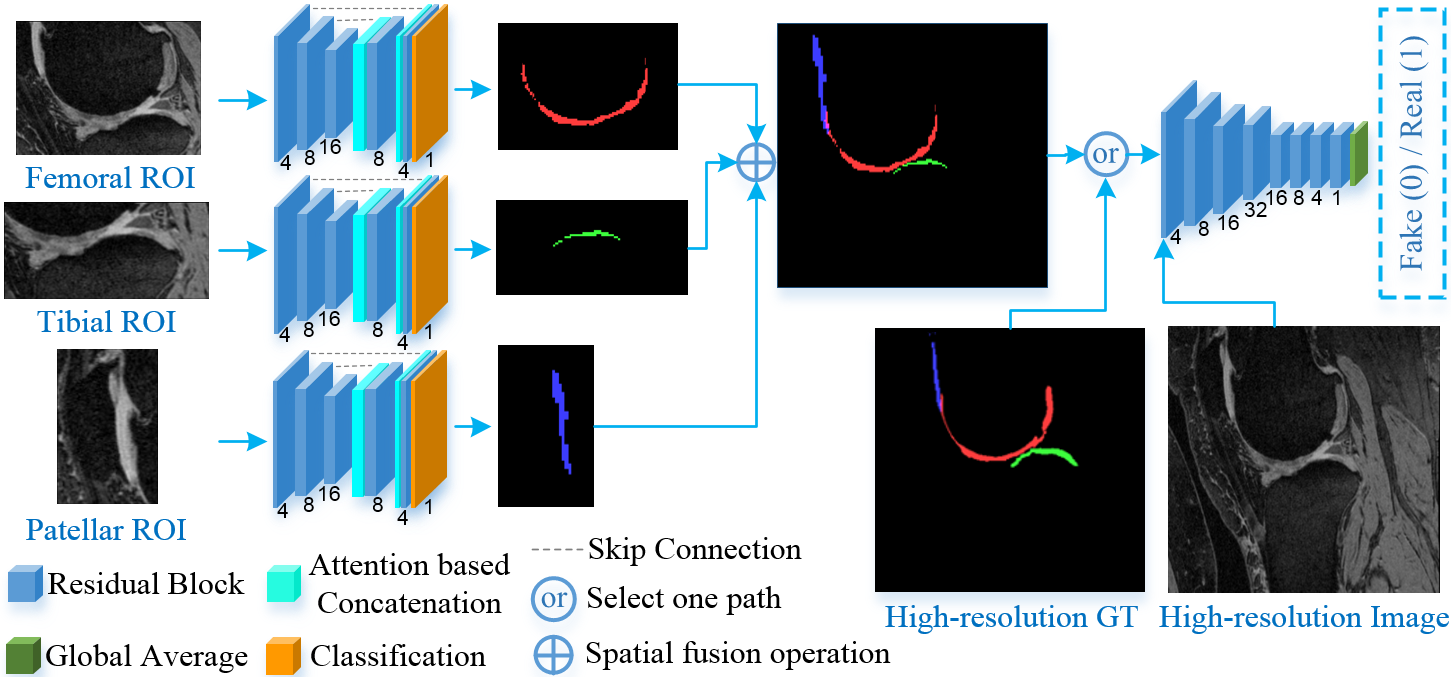}
    \caption{Demonstration of the collaborative multi-agent learning framework for fine-grained cartilage segmentation. The agents yield binary labels and the spatial fusion operation outputs a 4-channel result (FC, TC, PC and background).}
    \label{fig:critic}
\end{figure}

\noindent \textbf{Collaborative multi-agent learning}. In this learning stage (shown in Fig. \ref{fig:critic}), we construct one big network by three individual segmentation agents, one ROI-fusion layer, and one adversarial sub-network. The segmentation agent $A_{c = \left\{ {f,t,p} \right\}}$ ($f$, $t$ and $p$ stand for FC, TC and PC, respectively) aims to generate fine cartilage binary mask $A_c({\bf{x}}_{i, c})$ in the respective ROI ${\bf{x}}_{i, c}$ (its ground truth (GT) ROI is ${\bf{y}}_{i, c}$ and $i$ is the data index). Each ROI is small enough to cover only one cartilage in it. Since the large portion of background and other cartilages are excluded, the class imbalance problem is relieved significantly. The small ROIs also reduce the requirement for the computational resources (i.e., GPU memories) and enable fine-grained segmentation in high-resolution data. All the segmentation agents have similar VNet-like pattern as the coarse segmentor. To balance the receptive field of neurons and the GPU memory consumption, we further reduce the down- and up-sampling operations to 2. Considering the thin characteristics and unclear boundary of cartilage, we need to better utilize the multi-resolution contextual features to capture its fine details. In VNet, skip connection is designed to merge the up-sampled high-level features $I_h^{up}$ in decoding path and the equivalent-resolution low-level features $I_l$ in symmetrical encoding path by simple concatenation. Here, we apply an attention mechanism~\cite{jetley2018learn} to extend the skip connections. Formally, the connecting operation becomes $o\left( {\alpha \odot {I_l},I_h^{up}} \right)$, where $o$ denotes concatenation along the channel dimension, and $\odot$ is element-wise multiplication. The attention mask $\alpha = m\left( {{\sigma _r}\left( {{c_l}\left( {{I_l}} \right) + {c_h}\left( {I_h^{up}} \right)} \right)} \right)$ serves as a weight map that guides the learning to focus on desired region. Here, $c_h$ and $c_l$ are two convolutions of filter size $1$ and stride 1; $\sigma _r$ is an activation function (e.g., ReLU); $m$ is another convolution of filter size $1$ and stride 1 with sigmoid to contract the features to a single-channel mask. Light blue block in Fig. \ref{fig:critic} represents the novel attention based concatenation.

Although individual agent can obtain fine segmentation in its ROI, the individual learning losses the mutual constraints between cartilages. In order to make the agents collaborate together to make use of the mutual position and shape priors of all the cartilages for better delineations, we propose a collaborative learning strategy. This strategy utilizes a ROI-fusion layer $\cal F$ to restore the single-cartilage output from each agent back to the original knee joint space where the mutual constraints and priors can be encoded. ${\cal F}(A_f, A_t, A_p)$ is implemented by using the location information of the three input ROIs to fuse the fine cartilage masks back to the original space. Then, the multi-cartilage priors are learned implicitly by adversarial learning strategy. We utilize a discriminator sub-network $D$ to classify the fused multi-cartilage mask as ``fake'' and the whole GT label ${{\bf{y}}_i}$ as ``real''. In adversarial learning, the agents and the discriminator are trained alternatively. The parameters of agents are fixed when training the discriminator, and vice verse. In this way, discriminator sub-network can learn joint priors of multiple cartilages and guide the agents to produce better segmentation. It is important to note that the layer $\cal F$ not only fuses ROIs by their coordinates, but also passes the gradient updates from the discriminator to the agents during backpropagation, so that the two parts can be optimized in this alternating fashion. Since it is not intuitive to judge the labels without seeing the input in segmentation task, we borrow the idea of conditional generative adversarial nets, and treat the input MR knee image ${\bf{x}}_i$ as the conditioning variable. Fig. \ref{fig:critic} shows that the discriminator sub-network consists of 4 down-sampling convolutional layers, and the same residual block in the agents is also employed under each resolution level for contextual information learning. The input to the discriminator is a pair of MR knee image ${\bf{x}}_i$ and multi-label cartilage mask (either the GT label ${{\bf{y}}_i}$ or ${\cal F}(A_f, A_t, A_p)$). A global average layer is utilized at the end to generate a probability value for fake/real mask discrimination.

The loss functions of discriminator and agents are defined in Eq.~\ref{eq:cGAN_loss} and Eq.~\ref{eq:agents_loss}. Here, $\ell _b$ indicates the binary cross entropy loss. In Eq.~\ref{eq:agents_loss}, the first term ${L_s} = {\ell _b}\left[ {{A_c}\left( {{{\bf{x}}_{i,c}}} \right), {{\bf{y}}_{i,c}}} \right]$ is to train each single segmentation agent. The second term ${L_m} = {\ell _{mce}}\left[ {{\cal F}(A_f, A_t, A_p)}, {\bf{y}}_i \right]$ and the third one are applied on the fused multi-cartilage mask for joint-label learning. The discriminator $D$ and segmentation agents $A_{c = \left\{ {f,t,p} \right\}}$ are alternatively trained by minimizing Eq.~\ref{eq:cGAN_loss} and Eq.~\ref{eq:agents_loss}.

\begin{equation}\label{eq:cGAN_loss}
\sum\nolimits_i {\left\{ {{\ell _b}\left[ {D\left( {{{\bf{x}}_i},{{\bf{y}}_i}} \right),1} \right] + {\ell _b}\left[ {D\left( {{{\bf{x}}_i},{\cal F}\left( {{A_f},{A_t},{A_p}} \right)} \right),0} \right]} \right\}}
\end{equation}
\begin{equation}\label{eq:agents_loss}
\sum\nolimits_i {\left\{ {\sum\nolimits_{c = \left\{ {f,t,p} \right\}} {{L_s}\left( {{{\bf{x}}_{i,c}},{{\bf{y}}_{i,c}}} \right)}  + {L_m} + {\ell _b}\left[ {D\left( {{{\bf{x}}_i},{\cal F}\left( {{A_f},{A_t},{A_p}} \right)} \right),1} \right]} \right\}}
\end{equation}

\section{Experiments}
\noindent \textbf{Experimental settings}. We validate our proposed method on the iMorphics dataset from the OAI database. This set includes 176 3D MR (sagittal DESS sequences) knee images. The set is splitted into training: 120, validation: 26, testing: 30. Patients are randomly and exclusively used in the three subsets. Fixed ROI size of each type of cartilage is pre-defined based on adequate evaluation on the training data. We compare the proposed method with the state-of-the-art dense atrous spatial pyramid pooling (DenseASPP) for semantic segmentation~\cite{yang2018denseaspp}. It integrates the ASPP architecture in a dense connection manner, which is able to generate large receptive field and multi-scale features for segmentation tasks. We also evaluate performances of the proposed coarse segmentor and individual agents to show the effectiveness of the collaborative learning. Dice similarity coefficient (DSC), volumetric overlap error (VOE) and average surface distance (ASD) between the GT labels and segmented results are reported. In the training (no pre-trained weights used), we set the batch size to 1 and multiply a factor of 0.95 every 10 epochs to reduce the learning rate (LR). The Adam (with initial LR 0.001) and stochastic gradient descent (SGD, with initial LR 0.0002) solvers are used for each agent and the discriminator. All the networks are trained and tested by a 12GB-RAM Titan X GPU.

\begin{table}[!t]
  \caption{Quantitative comparisons of approaches: mean and std of evaluation metrics.}
  \label{tab:Quantification_Table_1}
  \centering
    \begin{tabular}{|l|c|c|c|c|c|c|c|c|c|c|c|c|}
    \hline
    {} & \multicolumn{3}{c|}{\textbf{Femoral Cartilage}} & \multicolumn{3}{c|}{\textbf{Tibial Cartilage}} & \multicolumn{3}{c|}{\textbf{Patellar Cartilage}} & \multicolumn{3}{c|}{\textbf{All Cartilages}} \\
    \cline{2-13}
    {} & \emph{DSC} & \emph{VOE} & \emph{ASD} & \emph{DSC} & \emph{VOE} & \emph{ASD} & \emph{DSC} & \emph{VOE} & \emph{ASD} & \emph{DSC} & \emph{VOE} & \emph{ASD} \\
    \hline
    $D1$ & 0.862 & 24.15 & 0.103 & 0.869 & 22.93 & 0.104 & 0.844 & 26.65 & 0.107 & 0.866 & 23.59 & 0.095 \\
    {} & 0.024 & 3.621 & 0.042 & 0.034 & 5.184 & 0.061 & 0.052 & 7.429 & 0.049 & 0.023 & 3.475 & 0.026 \\
    \hline
    $D2$ & 0.832 & 28.64 & 0.131 & 0.879 & 21.38 & 0.088 & 0.861 & 23.69 & 0.091 & 0.851 & 25.94 & 0.111 \\
    {} & 0.025 & 3.618 & 0.059 & 0.038 & 5.972 & 0.055 & 0.040 & 6.027 & 0.051 & 0.023 & 3.393 & 0.036 \\
    \hline
    $C0$ & 0.814 & 31.30 & 0.205 & 0.806 & 32.42 & 0.199 & 0.771 & 35.74 & 0.350 & 0.809 & 31.99 & 0.213 \\
    {} & 0.029 & 4.155 & 0.095 & 0.033 & 4.577 & 0.055 & 0.132 & 14.56 & 0.129 & 0.031 & 4.350 & 0.095 \\
    \hline
    \textbf{P1} & 0.868 & 23.19 & 0.108 & 0.854 & 25.17 & 0.126 & 0.824 & 28.78 & 0.201 & 0.862 & 24.24 & 0.110 \\
    {} & 0.023 & 3.514 & 0.067 & 0.029 & 4.173 & 0.059 & 0.104 & 12.45 & 0.439 & 0.023 & 3.457 & 0.048 \\
    \hline
    \textbf{P2} & \textbf{0.900} & \textbf{18.82} & \textbf{0.074} & \textbf{0.889} & \textbf{19.81} & \textbf{0.082} & \textbf{0.880} & \textbf{21.19} & \textbf{0.075} & \textbf{0.893} & \textbf{19.19} & \textbf{0.073} \\
    {} & 0.037 & 6.006 & 0.041 & 0.038 & 6.072 & 0.051 & 0.043 & 6.594 & 0.038 & 0.034 & 5.434 & 0.034 \\
    \hline
    \end{tabular}
\end{table}

\noindent \textbf{Experimental results}. Quantitative comparisons are shown in Table \ref{tab:Quantification_Table_1}. $C0$ represents the coarse cartilage extraction by the segmentor in Fig. \ref{fig:segmentor}. $\textbf{P1}$ denotes the fused results generated by the proposed segmentation agents, without the joint learning by the adversarial sub-network. $\textbf{P2}$ represents results from the proposed method by employing the collaborative multi-agent learning framework as in Fig. \ref{fig:critic}. For comparison, we integrate two variants of DenseASPP into the collaborative multi-agent framework. In the first variant $D1$, the residual blocks and skip connections are replaced by DenseASPP blocks in the two down-sampled levels of the agent network. While in the second variant $D2$, only the deepest level is replaced with DenseASPP block.

\begin{figure}[!t]
  \centering
  \begin{minipage}[b]{0.3\linewidth}
    \centering
    \centerline{\includegraphics[width=\linewidth]{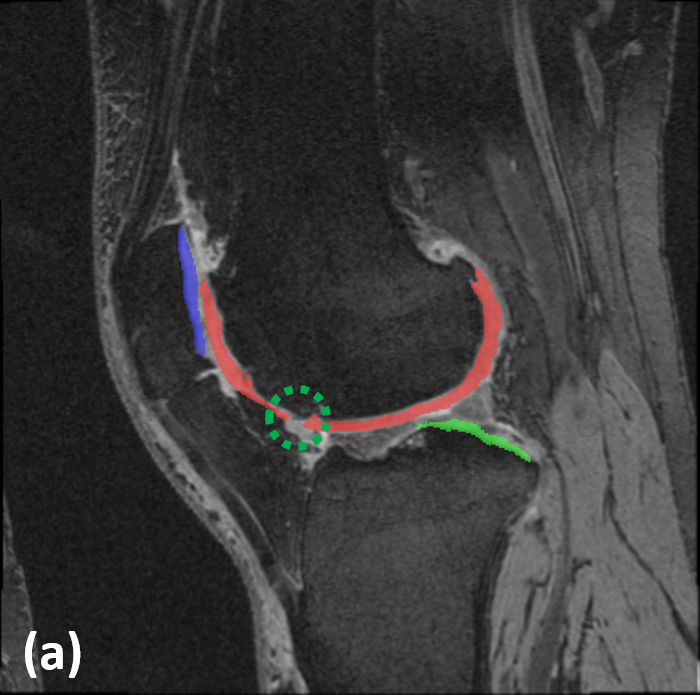}}
  \end{minipage}
  \hfil
  \begin{minipage}[b]{0.3\linewidth}
    \centering
    \centerline{\includegraphics[width=\linewidth]{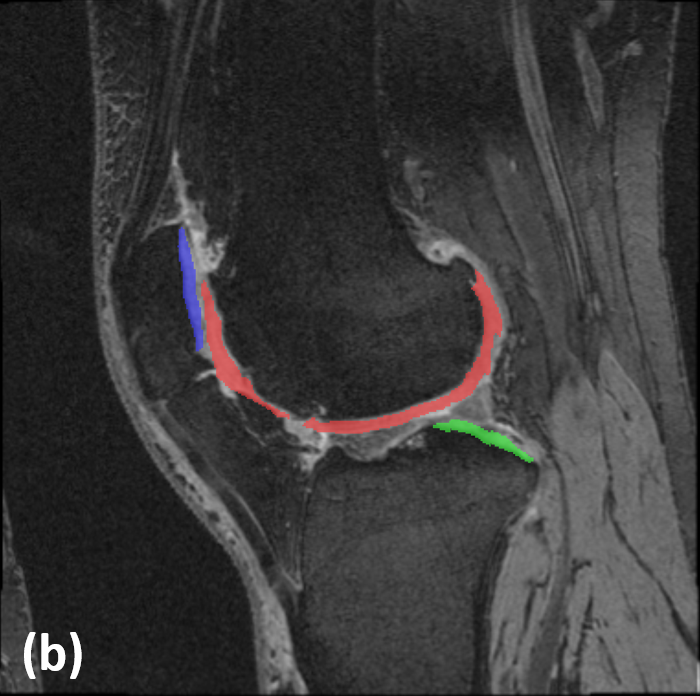}}
  \end{minipage}
  \hfil
  \begin{minipage}[b]{0.308\linewidth}
    \centering
    \centerline{\includegraphics[width=\linewidth]{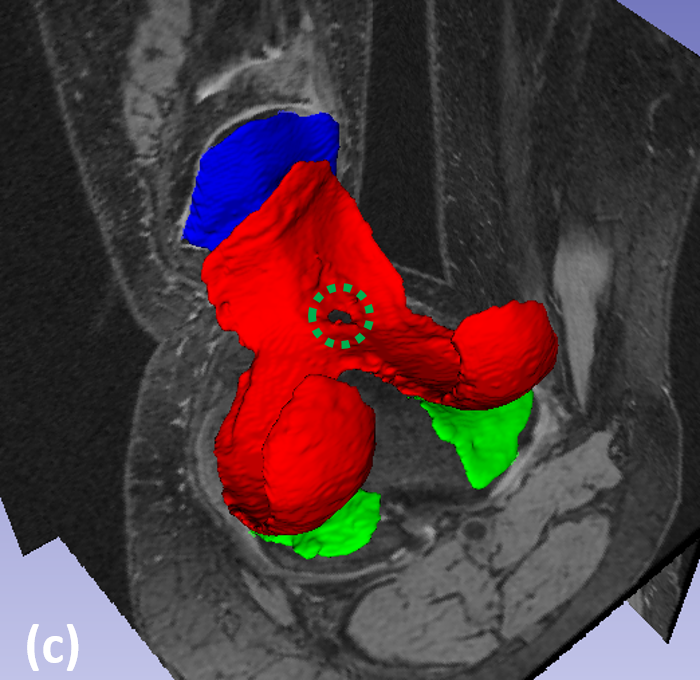}}
  \end{minipage}
  \caption{Results of subject 1. (a) and (b) show the segmentation and GT labels for FC (red), TC (green), and PC (blue) in sagittal view. (c) is the segmented 3D cartilages.}
  \label{fig:visual_oai_images}
\end{figure}

\begin{figure}[!t]
  \centering
  \begin{minipage}[b]{0.3\linewidth}
    \centering
    \centerline{\includegraphics[width=\linewidth]{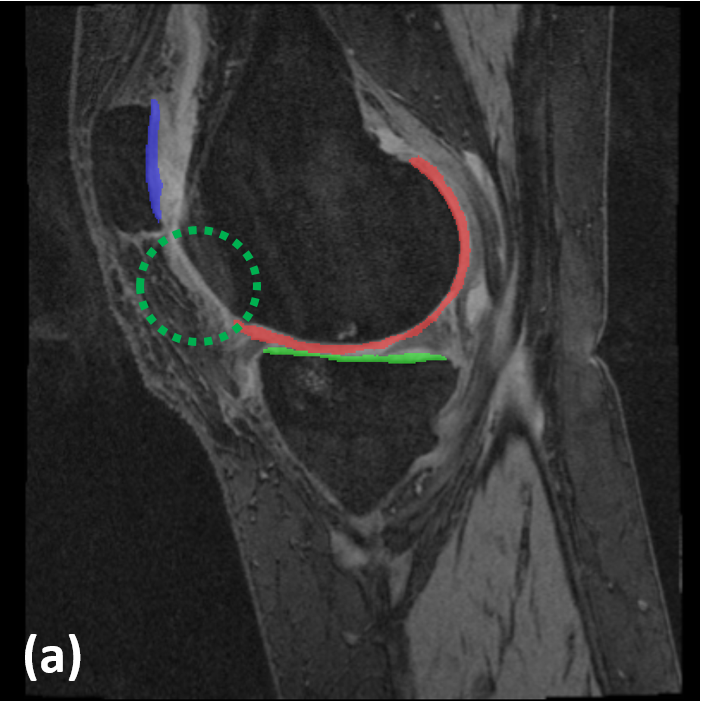}}
  \end{minipage}
  \hfil
  \begin{minipage}[b]{0.3\linewidth}
    \centering
    \centerline{\includegraphics[width=\linewidth]{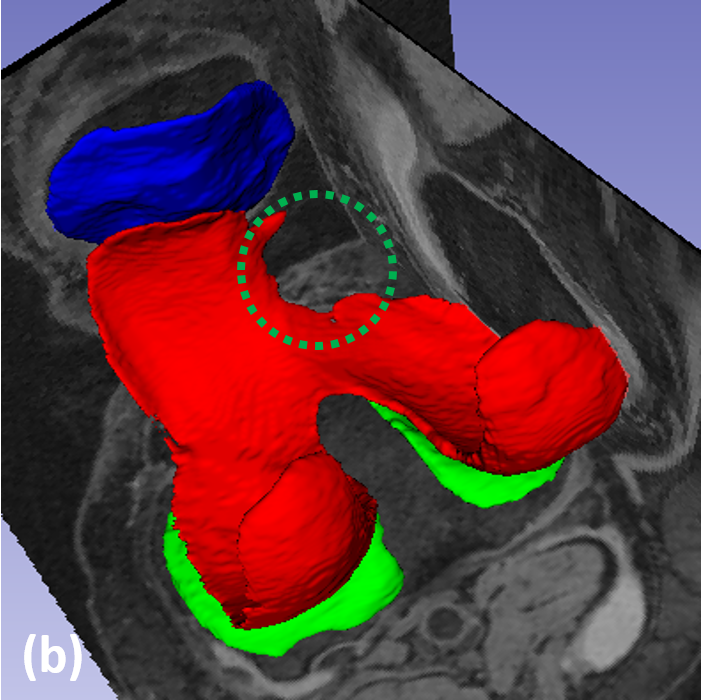}}
  \end{minipage}
  \hfil
  \begin{minipage}[b]{0.3\linewidth}
    \centering
    \centerline{\includegraphics[width=\linewidth]{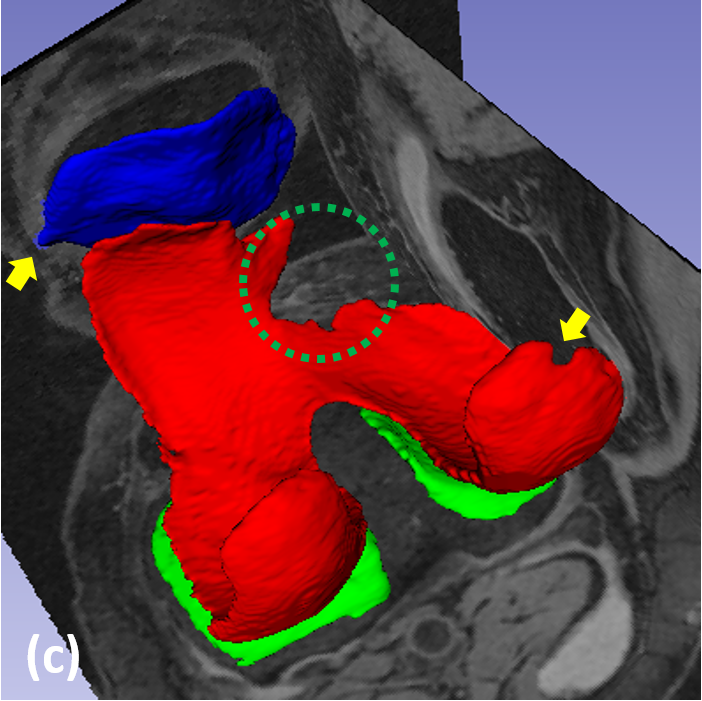}}
  \end{minipage}
  \caption{Results of subject 2. (a) shows the segmented cartilages in sagittal view. (b) and (c) demonstrate the GT and segmentation results in 3D view.}
  \label{fig:visual_oai_images_2}
\end{figure}

From the table, we can see that the proposed segmentation $\textbf{P2}$ achieves the best performance in all metrics. The mean results of $C0$ (i.e., a similar implementation of VNet) are relatively good and have no gross failure in our experiments. This shows that the coarse stage is reliable initialization. The proposed $\textbf{P2}$ obviously outperforming $\textbf{P1}$ shows that segmentation agents are improved with the help of the proposed collaborative learning strategy. The overall performances of the DenseASPP based variants $D1$ and $D2$ are close to that of $\textbf{P1}$. It indicates that the proposed agent network with the attention based concatenation is effective enough, compared to the DenseASPP blocks which have more complicated architecture. In addition, the results of the proposed method are comparable to those reported in some recent studies~\cite{ambellan2018automated,xu2018contextual}. Xu et al.~\cite{xu2018contextual} reported a total DSC $(0.887\pm0.024)$ value of FC and TC. Ambellan et al.~\cite{ambellan2018automated} utilized both 2D and 3D deep learning based segmentations with statistical shape models as shape refinement postprocessing for femoral and tibial cartilages extraction. Using a similar set from OAI, they achieved\footnote[2]{\cite{ambellan2018automated} separately presents the results of FC, medial TC and lateral TC at two timepoints. For convenience, we average these results and get the approximate mean/std metrics.} DSC $(0.893\pm0.024)$, VOE $(19.4\pm3.87)$ and ASD $(0.19\pm0.09)$ for FC, DSC $(0.881\pm0.038)$, VOE $(21.05\pm5.808)$ and ASD $(0.223\pm0.143)$ for TC. Without the sophisticated shape adjustment step, the proposed method acquires the comparable DSC and VOE scores, and much lower surface distance errors. Hence, the proposed framework can be used to automatically generate reliable assessments of all important articular cartilages in quantitative analysis for knee OA.

Visualization results (two examples) of the proposed method are showed in Fig.~\ref{fig:visual_oai_images} and \ref{fig:visual_oai_images_2}. The two patients have obvious shape variance of cartilages. In Fig.~\ref{fig:visual_oai_images} (a)-(c), the proposed method can accurately extract most of the cartilage regions and obtain smooth tissue boundaries. Furthermore, as indicated by green dashed circles in Fig.~\ref{fig:visual_oai_images} (a) and (c), our method can effectively capture a small cartilage defect. The green dashed circles in Fig.~\ref{fig:visual_oai_images_2} (a) and (c) indicate a possible cartilage damage/miss symptom well captured by our method. The 3D view exhibiting accurate 3D pattern of cartilage defects could be very useful in visual study of cartilage-related diseases. The yellow arrows in Fig.~\ref{fig:visual_oai_images_2} (c) show some minor errors occurred in some neighborhood areas due to unclear boundaries.

\section{Conclusions}
In this paper, we present a novel fully automatic method to segment three knee cartilages in 3D MR images based on a collaborative multi-agent learning architecture. Each segmentation agent depicts the high-resolution cartilage mask in its coarsely (but efficiently) located ROI. A novel skip connection by multi-resolution attention mechanism is introduced to enhance the feature extraction of target, while suppressing confusing information in neighborhood areas. Then, the depicted multiple ROIs are spatially fused into the original space to form a multi-cartilage label image for collaborative learning. The collaboration of agents is implemented by the novel ROI-fusion layer followed by an adversarial discriminator to ensure the shape and position constraints. Learning of the agents and discriminator are conducted in an alternating fashion. In our experiments, the proposed method achieves robust and accurate segmentation for all important articular cartilages in high resolution and large 3D MR knee data. In future we will apply the method for quantifing cartilage biomarkers (e.g., volume, thickness, surface area) in large-scale studies and detecting cartilage defects for lesion estimation~\cite{eckstein2010quantitative,hunter2011evolution}.
Besides the cartilages, the proposed framework could also be extended for other multi-organ segmentation tasks~\cite{uzunbacs2013collaborative}.

\bibliographystyle{splncs04}
\bibliography{paper807}
\end{document}